\documentclass[graybox]{svmult}

\usepackage{mathptmx}       
\usepackage{helvet}         
\usepackage{courier}        
\usepackage{type1cm}        
\usepackage{makeidx}         
\usepackage{graphicx}        
\usepackage{multicol}        
\usepackage[bottom]{footmisc}

\usepackage{amsmath,amssymb}
\usepackage{etoolbox}
\apptocmd{\thebibliography}{\raggedright}{}{}

\usepackage[utf8]{inputenc}
\usepackage{cite}

\usepackage{csquotes}

\makeindex             

\begin{document}

\title*{Database-driven High-Throughput Calculations and Machine Learning Models for Materials Design}
\titlerunning{High-Throughput and Machine Learning Models for Materials Design}
\author{Rickard Armiento}
\institute{Rickard Armiento \at Department of Physics, Chemistry and Biology, Link\"oping University, Sweden\\\email{rickard.armiento@liu.se}}
%
%
\maketitle

\abstract{This paper reviews past and ongoing efforts in using high-throughput \emph{ab-inito} calculations in combination with machine learning models for materials design. The primary focus is on bulk materials, \emph{i.e.}, materials with fixed, ordered, crystal structures, although the methods naturally extend into more complicated configurations. Efficient and robust computational methods, computational power, and reliable methods for automated database-driven high-throughput computation are combined to produce high-quality data sets. This data can be used to train machine learning models for predicting the stability of bulk materials and their properties. The underlying computational methods and the tools for automated calculations are discussed in some detail. Various machine learning models and, in particular, descriptors for general use in materials design are also covered.}
  
\section{Background}
\label{sec:1}
Design of new materials with desired properties is a crucial step in making many innovative technologies viable. The aim is to find materials that fulfill requirements on efficiency, cost, environmental impact, length of life, safety, and other properties. During the past decades, we have seen major progress in theoretical materials science due to the combination of improved computational methods and a massive increase in available computational power. It is now standard practice to obtain insights into the physics of materials by using supercomputers to find numerical solutions to the basic equations of quantum mechanics. When using the appropriate level of theory, these calculations can be robust enough to run in unsupervised high-throughput. Hence, materials design can be done via automated theoretical screening of candidate materials and substances, picking out those with desired properties. Early examples of this methodology include works in the fields of catalysts \cite{greeley_computational_2006}, battery materials \cite{kang_electrodes_2006, kirklin_high-throughput_2013}, detector materials for ionizing radiation \cite{ortiz_data_2009}, superconductivity \cite{klintenberg_possible_2013}, thermoelectricity \cite{madsen_automated_2006,wang_assessing_2011}, piezoelectrics \cite{armiento_screening_2011,armiento_high-throughput_2014}, transparent conducting oxides \cite{hautier_identification_2013}, and two-dimensional materials \cite{lebegue_two-dimensional_2013}. There is a wealth of further examples in the literature, see \emph{e.g.}, the reviews in Ref.~\cite{curtarolo_high-throughput_2013,ceder_how_2013,alberi_2019_2019,oba_design_2018}.

Early adoption of high-throughput methodology for materials design have invoked the ambition that it may be possible to computationally predict the existence and basic properties of essentially \emph{every single material}, \emph{i.e.,} any composition that, in principle, can be synthesized as a reasonably long-lived `stable' compound (in the context of an environment.) This ambition has been referred to as the \emph{materials genome project} \cite{ceder_how_2013, jain_high-throughput_2011, jain_commentary:_2013}, which in 2011 was endorsed as a White House initiative \cite{executive_office_of_the_president_national_science_and_technology_council_washington_dc_materials_2011}. The idea is that access to materials genome data with sufficient coverage would greatly accelerate materials design. It would be possible to perform queries against this data to pick out compositions that have some sought combination of desired properties for a specific application at, essentially, no additional computational cost \cite{ceder_how_2013, curtarolo_high-throughput_2013}.

A large number of databases of materials-genome-type are now available, many of them open and free for access over the Internet. Some notable examples include: the \emph{Electronic Structure Project} (http://gurka.physics.uu.se/esp/; 2002), the \emph{Automatic FLOW repository} (aflowlib.org; 2011), \emph{the Materials Project} (materialsproject.org; 2011), the \emph{Open Materials Database} (openmaterialsdb.se; 2013), the \emph{Open Quantum Materials Database} (oqmd.org; 2013), the \emph{Theoretical Crystallographic Open Database} (www.crystallography.net/tcod; 2013), the \emph{Novel Materials Discovery Repository} (nomad-repository.eu; 2014), the \emph{High Performance Computing Center Materials Database - NREL MatDb} (materials.nrel.gov; 2015), and the \emph{Materials Cloud} (materialscloud.org; 2017).

To use machine learning models for, \emph{e.g.}, molecular dynamics simulations of systems with up to a few chemical species has become increasingly popular (\emph{i.e.}, to accelerate simulations of the movement of some types of atoms in a material.) To train more general models with data from materials genome-type databases opens a way forward towards the vision of a complete coverage of materials and their properties. This paper reviews the use of high-throughput techniques and tools to produce training data for these models and recent developments in the area of models with the aim of a general description of atomistic systems (\emph{i.e.}, molecules and materials.) This development is, at its core, the adoption of an informatics perspective to materials science and design, which has been referred to as \emph{materials informatics} \cite{rajan_materials_2005, rodgers_materials_2006}.

It has been posed as a hypothesis that the progress of general AI methods will eventually reach `the singularity', a moment in time when self-improving AI methods set off a runaway technological development that fundamentally changes society\footnote{The term was recently popularized by a 2006 book by Kurzweil \cite{kurzweil_singularity_2006}, but its use goes back to a 1958 account by Stanislaw Ulam of a discussion with John von Neumann that references a point in time of fundamental change due to runaway technological development \cite{ulam_john_1958}.}. One can, in a similar way, formulate the hypothesis that the development of increasingly sophisticated machine learning models for atomistic systems will reach a singularity of its own, \emph{i.e.,} a point in time of fundamental change in our theoretical description of physical matter. This change would happen when fully trained, general, machine learning models appear that are capable of predictions at the same accuracy as physics-based quantum mechanical simulations but at negligible computational effort. The result would turn the present materials-genome type databases obsolete and enable true inverse design of molecules and materials with desired properties across the full chemical space at near zero computational expense. Such a development would bring far-reaching changes across the natural sciences.

In conclusion, advancing the present state of materials design with machine learning models requires progress in three key areas: \emph{(i)} progress in the theory and methods used in physics-based calculations that can be used to improve the quality of training data. This requires developing methods with improved accuracy without sacrificing the low computational demand and the high level of generality that are necessary for the methods to be useful for high-throughput calculations; \emph{(ii)} further improved methods and tools for running automated calculations in high throughput. While there are many software packages and solutions available today for running calculations in high-throughput, major work of both practical and theoretical nature remains to turn methods that were developed and tested only on a few systems into automated workflows capable of running unsupervised at large scale without human interference; and \emph{(iii)} further improved machine learning models for general atomistic systems. 

\section{Computational methods}

Kohn-Sham density functional theory \cite{hohenberg_inhomogeneous_1964,kohn_self-consistent_1965} (KS-DFT) is the standard theoretical framework for high-throughput computation in present materials property databases. There is a range of software implementations for performing the numerical solution of the basic equations of DFT. A few prominent examples include the \emph{Vienna Ab-inito Simulation Package - VASP} (vasp.at), \emph{ABINIT} (www.abinit.org), \emph{Wien2K} (susi.theochem.tuwien.ac.at), and \emph{Quantum ESPRESSO} (www.quantum-espresso.org). Of primary concern for these software packages is the numerical convergence towards the exact solution with respect to the approximations used. Most approximations are fairly straightforward to systematically improve towards a converged result, which has led to a number of standard practices for setting convergence parameters that are typically documented in relation to the respective database. See, \emph{e.g.}~Ref.~\cite{jain_commentary:_2013} for the practices used in the Materials Project database.

One aspect of numerical convergence that frequently is in focus when discussing the accuracy of KS-DFT calculations in the context of chemistry-oriented calculations is the basis set used to represent the single particle wave functions (also known as the \emph{KS orbitals.}) While more or less all basis sets can be systematically extended towards numerical convergence, this can be impractical for some choices. Nevertheless, in the context of materials design of bulk materials, we are mostly concerned with fully periodic crystals where the most common choice is a plane-wave basis set where systematic convergence is more straightforward.

In contrast to the numerical approximations that can be, at least in principle, systematically refined to arbitrary accuracy towards the solution of the KS-DFT equations, there is one aspect of the calculations where this is not possible. This is the choice of exchange-correlation density functional. This choice is crucial for the description of the physics of the system and, by extension, which properties are available in the output. The kind of systems and properties for which one can obtain reliable data is of key importance in the present context of using high-throughput computation to produce reliable training data for machine learning models. Hence, we will in the following review the important aspects of this choice in detail.

The level of theory that so far has been the standard for high-throughput computation in first-principles materials property databases is the semi-local, `second-rung' \cite{perdew_jacobs_2001} level, which uses exchange-correlation functionals on the generalized gradient approximation (GGA) form. The most commonly used functional in the context of high-throughput calculations for materials databases is the one by Perdew, Burke and Ernzerhof \cite{perdew_generalized_1996} (PBE) with the $+U$ correction \cite{anisimov_first-principles_1997}. This level of theory strikes a desirable balance between computational speed and accuracy while maintaining a high level of transferability. Nevertheless, the most popular GGA-type functionals, including PBE, have known shortcomings in their description of the electronic structure. The primary issues include: \emph{(i)} a tendency to give energetics that in geometrical relaxations lead to a systematic over- or underestimation of bond lengths (the local density approximation, LDA, overbinds, whereas PBE underbinds); \emph{(ii)} an insufficient description of the physics of weak dispersion forces/van der Waals bonding; and \emph{(iii)} a systematically overdelocalized description of the KS orbitals that leads to inaccuracies in a number of properties that are derived from the orbitals. These theree issues will be discussed in some more detail in the subsections below.

\subsection{Overdelocalized orbitals}
The fundamental issue of overdelocalized KS orbitals is related to various aspects of the self-interaction error present at the semi-local exchange-correlation functional level of theory. A simplified picture is that the self-interaction introduces a repulive electrostatic interaction of an electron with itself, leading to a delocalization that becomes more severe the more localized the correct representation of the orbital was supposed to be, \emph{i.e.,} the effect more severely impacts the more localized $d$-, and even more so, the $f$ orbitals, compared to the less localized $s$ and $p$ states. The result is a number of deficiencies in predicted materials properties. Examples of problematic properties include redox reaction energies \cite{zhou_first-principles_2004, chevrier_hybrid_2010}, the polarizability of extended systems \cite{kummel_electrical_2004, armiento_electrical_2008}, and the silicon interstitial formation energy \cite{mattsson_electronic_2008, rinke_defect_2009}.

In addition to these examples, issues are also seen in a number of properties calculated from the single-particle orbitals from the KS-DFT framework, where they are used as approximations of the `true quasi-electron orbitals' of the many-electron system (to the extent that such can be defined). However, from a fundamental perspective, the discussion of the accuracy of such properties is delicate because the DFT orbitals and the quasi-electron orbitals are not the same thing, even in theory for the exact exchange-correlation functional. Hence, one cannot \emph{a priori} assume that an improved functional increases the agreement with the experimental values of, \emph{e.g.,} optical properties calculated from the KS band structure. Nevertheless, if one compares common GGA functionals to higher order methods that are still within the framework of KS-DFT (\emph{e.g.,} exact DFT exchange) one finds a qualitative difference in the orbital physics. This difference translates to that when materials properties which are directly associated with the electronic structure are calculated using higher-order theory, the results come out qualitatively closer to experiments than those calculated using standard GGAs. One can, therefore, take the position that it is a worthwhile improvement over standard semi-local functionals if improved functionals can make the orbitals to more closely mimic the orbital features given by higher-order methods. This motivation is independent of the justification, or lack thereof, of using KS states to approximate quasi-particle bands for calculating materials properties. For an expanded discussion on this delicate topic, see, \emph{e.g.}, Ref.~\cite{vlcek_improved_2015}.

There are a range of well-known methods to address the description of localized states in semi-local DFT, \emph{(i)} an explicit orbital-dependent correction that removes the surplus electrostatic term (sic correction) \cite{perdew_orbital_1979, perdew_self-interaction_1981, jones_density_1989}; \emph{(ii)} exact exchange DFT \cite{stadele_exact_1999}; \emph{(iii)} interpolating the DFT functional with Hartree-Fock exchange energy (hybrid functionals) \cite{becke_densityfunctional_1993, heyd_hybrid_2003, heyd_erratum:_2006}; \emph{(iv)} use of the many-body Green's function for a more precise description of the localized quasi-particle orbitals (GW) \cite{hedin_new_1965}; \emph{(v)} the DFT$+U$ correction that adds an effective Hubbard-like term to the Hamiltonian to make selected localized orbitals energetically preferable \cite{anisimov_first-principles_1997}; and \emph{(vi)} various attempts to modify the KS potential directly to make it reproduce essential features of exact exchange \cite{armiento_electrical_2008, van_leeuwen_exchange-correlation_1994, gritsenko_self-consistent_1995, becke_simple_2006, umezawa_explicit_2006, rasanen_universal_2010, tran_accurate_2009}. All these methods, except for the last two (v, vi), require a vastly increased computational expense. Hybrid density functional methods (iii) are increasingly adopted for resolving these issues when the extra computational cost is acceptable. However, at a cost of roughly 50 times of that of standard GGAs, they are very inconvenient, or even completely unsuitable for, \emph{e.g.}, larger systems and high-throughput-type calculations.

Of the two less computationally expensive methods, DFT$+U$ (v) is widely adopted as, arguably, the standard way of dealing with the issue of overdelocalized orbitals in high-throughput calculations and materials genome-type databases. However, DFT$+U$ is not a highly transferable method; it requires attention in the assignment of site-specific ``$U$-values''. In setting the value of $U$, one selects how strongly a given localized orbital on a specific site prefers full occupation over partial occupation. In low throughput calculations, it is common to somewhat thoroughly investigate a system to arrive at a value of $U$ that reasonably reproduces the expected physics of the system, but this is clearly not an option for high-throughput calculations. There are schemes to obtain sets of values that work well for systems with some specific type of physics, \emph{e.g.,} for typical oxides. However, in systems of mixed chemistries and intermixed types of bonding physics, the non-universality of $U$ values becomes a serious problem. Energies obtained for different systems using different $U$-values for the same species cannot easily be mixed. Furthermore, since $U$ values are usually only assigned to specific orbital projections on a pre-selected set of transition metal species, they cannot help with overdelocalized states of different origin, \emph{e.g.,} for defect states that are not atomic-orbital--like.

The second computationally less expensive method in the list above is (v) the approach to model the exchange-correlation potential directly to make it reproduce essential ``non-local'' features of exact exchange, instead of obtaining it as a functional derivative of an energy functional. Such potentials are known as \emph{model potentials,} and have in some cases been quite successful \cite{van_leeuwen_exchange-correlation_1994, gritsenko_self-consistent_1995, becke_simple_2006, umezawa_explicit_2006, armiento_electrical_2008, rasanen_universal_2010, tran_accurate_2009}. Some recent interest has been generated by the model potential of Becke and Johnson (BJ) \cite{becke_simple_2006}, which was observed to mimic some of the crucial features of exact exchange for atoms. With various adjustments and extensions, it improves the polarization of hydrogen chains \cite{armiento_electrical_2008, rasanen_universal_2010}, give closer correspondence to experimental band gaps \cite{tran_accurate_2009}, and, to some extent, gives other improved properties \cite{oliveira_toward_2010, singh_electronic_2010}. These model potentials seem promising for future adoption in high-throughput calculations to access properties that would otherwise be problematic because of orbital delocalization.

However, there are some fundamental issues with the general approach of model potentials. Since they directly model the exchange-correlation potential, the corresponding energy functionals are not merely unknown, they usually do not \emph{exist} \cite{umezawa_explicit_2006, van_leeuwen_energy_1995, gaiduk_construction_2011}, and this deficiency cannot easily be corrected \cite{karolewski_polarizabilities_2009}. Since the KS equations are derived from a variational treatment of an energy equation, the use of such potentials has to be regarded on a weak formal basis, and are, strictly, outside the framework of KS-DFT. One cannot calculate any energy-derived properties from model potentials, \emph{e.g.,} one cannot do a geometry optimization that is consistent with the potential. Hence, if one starts from, \emph{e.g.,} theoretically generated structure candidates, one would have to use another method first to pre-relax the structure.

A closely related promising direction of functional development is the Armiento-K{\"u}mmel exchange functional (AK13) \cite{armiento_orbital_2013} (co-authored by the author of this paper.) This is \emph{a normal GGA exchange energy functional} that mimics the behavior of the BJ potential while avoiding the fundamental issues with model potentials. Similar to the modified BJ-based model potentials, the AK13 exchange energy functional gives qualitatively different orbitals from common GGA functionals. The results are a KS potential with improved atomic shell structure \cite{armiento_orbital_2013}, improved ionization potentials from the highest eigenvalue \cite{armiento_orbital_2013} (but see the discussion in Ref.~\cite{aschebrock_challenges_2017}), overall a KS band structure that better match that of higher order methods, including enlarged band gaps, and improved optical properties \cite{armiento_orbital_2013,vlcek_improved_2015,cerqueira_benchmarking_2014,tran_comparison_2015}. As mentioned, the AK13 functional avoids the problem of undefined energies and energetics in model potentials. However, their values are not as accurate as those of commonly used GGAs and mostly insufficient. In addition, other issues appear from the AK13 construction that prevent its broader indiscriminate application \cite{lindmaa_energetics_2016,aschebrock_orbital_2017,aschebrock_challenges_2017}. We are hopeful that further research into modifications of the expression can overcome the difficulties while still retaining the favorable exchange potential features.

\subsection{Under- and overestimated lattice constants}
On the issue of systematic under- and overestimation of lattice constants, this has mostly been resolved in functional development beyond PBE. The Armiento-Mattsson 2005 functional (AM05) \cite{armiento_functional_2005, mattsson_implementing_2009} is a semi-local functional with the same computational difficulity as PBE, but which gives roughly half the error for lattice constants. The comprehensive testing of Haas \emph{et al.} finds for the lattice constants of 60 solids that the mean absolute error is 0.053~\AA\ for PBE and 0.033~\AA\ for AM05 \cite{mattsson_am05_2008,haas_calculation_2009, haas_erratum:_2009}. Later functionals developed by Wu-Chohen in 2006 \cite{wu_more_2006,zhao_comment_2008,wu_reply_2008}, SOGGA by Zaho, and Truhlar in 2008 \cite{zhao_construction_2008}, and PBEsol by Perdew \emph{et al.} in 2009 \cite{perdew_restoring_2008,mattsson_comment_2008,perdew_perdew_2008,perdew_erratum:_2009} report similar improvements \cite{mattsson_comment_2008, haas_calculation_2009, haas_erratum:_2009}. Further progress have been made by Perdew and coworkers on the meta-GGA level of theory, where, in addition to the electron density and its derivatives, a functional may also depend on the local value of the kinetic energy density of the KS particles. While meta-GGAs are technically more complex expressions that GGAs, implementations can be made that do not significantly increase the computational cost. The 2015 \emph{Strongly Constrained and Appropriately Normed Semilocal Density Functional} (SCAN) meta-GGA \cite{sun_strongly_2015} reportedly performs well for a wide range of properties for both solids and molecules, including lattice constants \cite{sun_accurate_2016,zhang_efficient_2018}. However, some issues have recently been reported in the description of systems with itinerant magnetism \cite{ekholm_assessing_2018}.

\subsection{Weak dispersion forces}
On the topic of the description of van der Waals/London dispersion forces/weak interactions by semi-local DFT functionals, there exist a range of post-correction schemes of the energy to handle such interactions that can be deployed without any significant additional computational cost, see \emph{e.g.}, Refs.~\cite{grimme_semiempirical_2006, grimme_consistent_2010, tkatchenko_accurate_2009,tkatchenko_accurate_2012,ambrosetti_long-range_2014,steinmann_comprehensive_2011,steinmann_generalized-gradient_2011}. Furthermore, there is a series of successful exchange-correlation functionals known as the vdW-DF from a collaboration between Chalmers University and Rutgers University \cite{dion_van_2004,lee_higher-accuracy_2010,berland_exchange_2014} which allow a self-consistent treatment of these interactions. These functionals are not semi-local, but still fairly computationally inexpensive compared to, \emph{e.g.,} hybrid functionals. Furthermore, it has been shown that information about weak interactions can be extracted from local values of the kinetic energy density which are available to meta-GGAs \cite{sun_strongly_2015,sun_density_2013}, at least to a level where the region around the equilibrium in van der Waals bonds can be described. This development has been incorporated in the SCAN functional \cite{sun_strongly_2015}. 

\section{Materials Properties}\label{sec:materials_properties}
One of the central questions with the materials-genome effort is what basic properties are within reach to be collected and included in these databases. This is determined by a combination of what can be described by the level of theory used for the computations (as carefully reviewed in the previous section), and what methods are available as automated workflows. The starting point, crucial for building any materials-genome type resource, is the crystal structures and corresponding formation energies. The importance of the formation energies is due to their use in creating composition phase diagrams to estimate the zero temperature thermodynamic stability of a material. The composition phase diagram gives the ground state crystal structure of a material at zero temperature as a function of composition. It is constructed by determining the convex hull of the predicted formation energies of all competing crystal structures in a chemical subspace \cite{akbarzadeh_first-principles_2007, ong_li-fe-p-o2_2008, jain_high-throughput_2011}. A compound with a formation energy on the convex hull is stable, whereas a compound that ends up above the hull is unstable. The distance to the hull can be used as a rough estimate of the degree to which a material is unstable (\emph{i.e.}, how unlikely it is to be observed, and if observed, how quickly it would deteriorate into a combination of lower energy structures.) Crystal structures with a small hull distance (very roughly up to $\sim$ 50 meV) may still be regarded as candidates for materials that in practice may be stable since such an `error margin' can account for meta-stability, stability at limited elevated temperatures, and the computational inaccuracy of the methods.

Several works have investigated the accuracy of DFT calculations of formation energies. The standard deviation of formation energies calculated with PBE$+U$ to experiments for the formation of ternary oxides from binary oxides was found to be 0.024 eV/atom; meaning 90\% of the errors are within 0.047\ eV/atom, which corresponds to a mean absolute error of approximately 0.02 eV/atom \cite{hautier_accuracy_2012}. Kirklin \emph{et al.} determined a mean absolute error of PBE formation energies of systems over all chemistries to be 0.136 eV/atom, but with energy corrections that are often used in high-throughput databases to some of the elemental phase energies, this lowers to 0.081 eV/atom \cite{kirklin_open_2015}. However, the same paper notes that for 75 intermetallic structures they found experimental results from more than one source, giving an estimate for the mean absolute error in the experiments of 0.082 eV/atom. (Note that the latter estimate may be affected by selection bias, \emph{i.e.}, there may be a larger probability of finding multiple experimental values if the results are uncertain.) 

Presently the set of materials properties beyond stability and formation energies available for large data sets is somewhat limited. There is an ongoing competition between the online materials genome-type databases to grow the data they provide both in terms of included structures and materials properties. While there is a wealth of published methods to produce data for a wide range of properties available in the literature, to turn methods that have been published into a form where they can run reliably in high-throughput is non-trivial. Among the available databases, the Materials Project is quite comprehensive in terms of properties. In addition to structural information and formation energies, they have over the years added the KS-DFT band structure (in some cases corrected using the GW approximation \cite{hedin_new_1965,castelli_new_2015}), elastic tensors \cite{de_jong_charting_2015}, piezoelectric tensors \cite{de_jong_database_2015}, dielectric properties \cite{petousis_high-throughput_2017}, phonon spectra \cite{petretto_high-throughput_2018}, synthesis descriptors \cite{kim_materials_2017}, and xray absorption spectra \cite{mathew_high-throughput_2018}.

\section{Database-driven High-Throughput Calculations}
\begin{figure}[t]
  \includegraphics[width=0.8\linewidth]{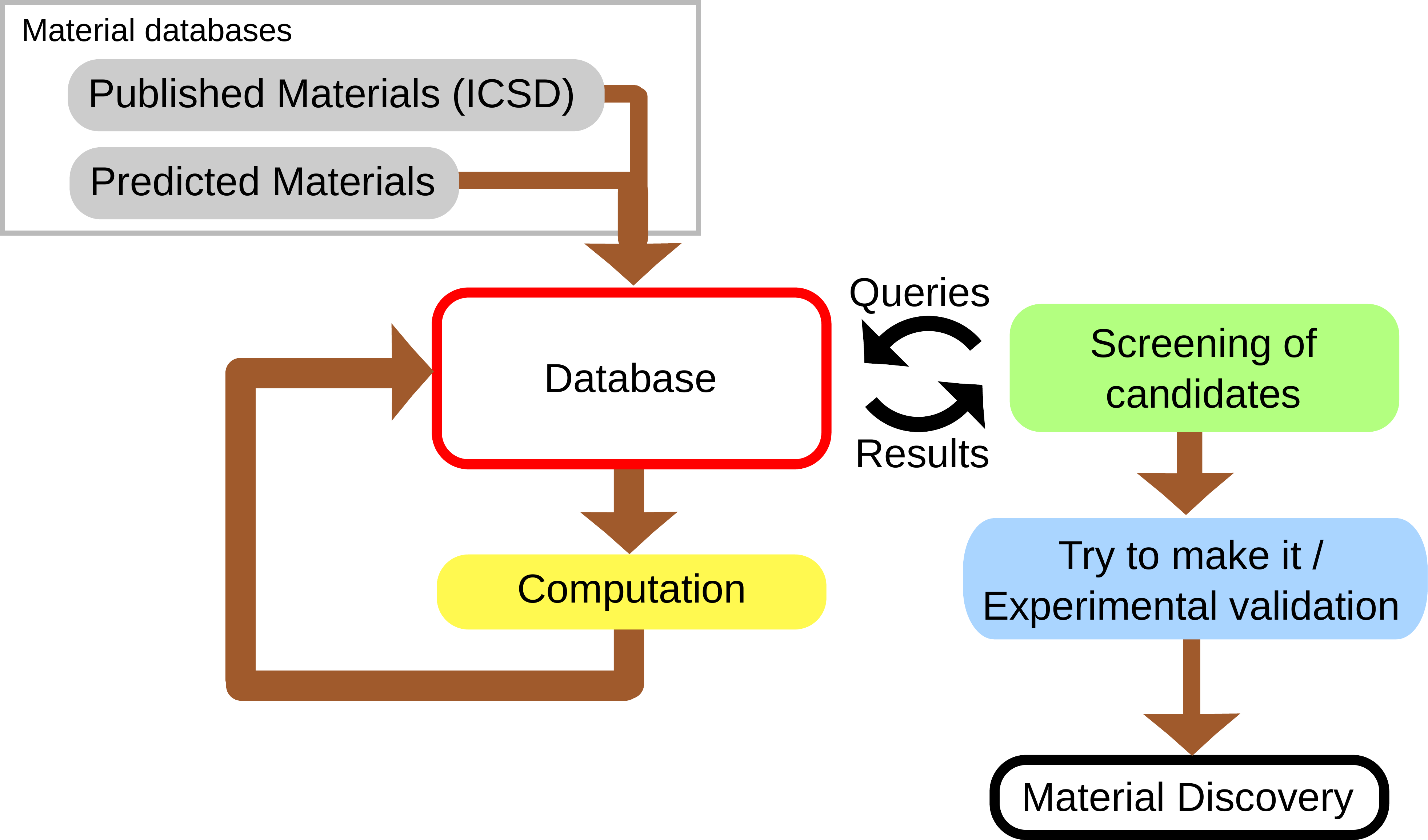}
\caption{A schematic flowchart representation of database-driven high-throughput materials design, largely inspired by the setup used in the Materials Project \cite{jain_high-throughput_2011}. The steps on the right hand sides represent the use of the database to find materials with desirable properties. In the context of machine learning models, the materials and materials properties in the database can be used for training and validation.\label{fig:flowchart}}
\end{figure}

A basic flowchart for materials design using database-driven high-throughput calculations is shown in Fig.~\ref{fig:flowchart}. There are many software packages with partially overlapping aims for helping with the steps in the flowchart. Some recognized open source examples are the \emph{atomic simulation environment - ASE} (wiki.fysik.dtu.dk/ase), \emph{pymatgen, custodian, and fireworks} (pymatgen.org, see also the information at materialsproject.org/infrastructure; connected to the Materials Project), \emph{aflow} (materials.duke.edu/AFLOW; connected to the AFLOW repository), \emph{AiiDA} (aiida.net; connected to materials cloud), \emph{qmpy} (connected to the open quantum materials database). The author is involved in the development of the open source \emph{high-throughput toolkit - httk} (httk.org) framework, which we use extensively for high-throughput computation in our own research, and which provides the backend for the open materials database. This toolkit provides functionality for preparing and running unsupervised workflows of calculations (electronic structure, mostly targeted towards the software package VASP), analyzing the results, and store them in a global and/or in a personalized database. The basic functionalities of these software packages are quite similar; in the following, we discuss the functionality of \emph{httk}.

The primary focus of \emph{httk} is for running automated calculations with as little human intervention as possible. This is crucial when working with large datasets, but can also be convenient when working with smaller projects. The toolkit consists of a software library developed in Python and a set of script programs that enable the interaction with supercomputers. The primary strengths of this framework compared to common alternatives is \emph{(i)} the Python library provides a very integrated object-relational mapper, where classes in object-oriented Python are used to introduce abstractions that remove much of the difficulty in setting up a personal database of SQL type in which one can store, search, retrieve, and analyze results; \emph{(ii)} \emph{httk} consistently allows the use of exact rational numbers in place of the more commonly used floating-point numbers. The exact rational numbers allows processing of crystal structures, application of transforms, etc., without the usual loss of precision. Hence, \emph{httk} can deterministically produce an internal representation of structures read from a source file (\emph{e.g.}, on the cif file format), which is not the case in most other frameworks due to their use of floating-point numbers means the precise end result is influenced by the computer architecture.

The \emph{httk} framework is distributed in several ways, including the PyPi service. Hence, it can easily be installed by issuing: \texttt{pip install httk} on a system with a modern distribution of Python. There is a set of tutorial steps and a large number of examples available to show how the framework can be utilized in the various steps of database-driven high-throughput as shown in Fig.~\ref{fig:flowchart}. These are available via the project website (httk.openmaterialsdb.se).

\section{Machine Learning Models for Materials Design}

\subsection{Models for molecules}
The primary focus in this paper is on a type of machine learning models for use in materials design that can be said to begin with a 2012 paper by Rupp \emph{et al.} on the use of kernel ridge regression for small molecules \cite{rupp_fast_2012}. They define a matrix representation for molecules named the `Coulomb matrix'. In this representation a system of $N$ atoms generates an $NxN$ matrix where the off-diagonal elements ($i$,$j$) are the Coulomb repulsion between the $i$th and $j$th bare atomic cores, and the diagonal elements are based on a polynomial fit to energies of free atoms to represent a static energy contribution pertaining to the $i$th atom,
\begin{eqnarray}
\label{ColumbMatrix}
C_{ij} &=&
  \begin{cases}
    0.5 Z_{i}^{2.4}   &\mbox{if } i = j  \\
    Z_{i}Z_{j} / (\|\mathbf{r}_{i}-\mathbf{r}_{j}\|_2) &\mbox{if } i\neq j 
  \end{cases}
\end{eqnarray}
One may note that the Coulomb interaction between the bare atomic cores is not a good indicator of the physics of the bonds in a system. However, the representation does not aim to push the machine learning model into a specific physics-based description, but just to constitute a well-formed way to represent the structural information (\emph{i.e.}, the positions of the atoms) so that the machine is free to learn the physics from the data. This model was trained on small organic molecules (with up to 7 atoms from the elements C, O, N, and S, and with the valencies satisfied by hydrogen atoms; this is data set is named \emph{qm7.}) It was shown in the original paper that the machine can be trained to predict atomization energies of molecules not in the training set down to a mean absolute error of 10 kcal/mol at a training set size of 7k. In units more common for materials, this model reaches $20\ \mathrm{meV/atom}$ at a training set of 3000 molecules from qm7 \cite{faber_crystal_2015}. 

\subsection{General models for periodic systems}
In a 2015 work Faber, Lindmaa, von Lilienfeld, and Armiento (the autor of the present paper) extended the Coulomb matrix construct into a suitable form for periodic crystals \cite{faber_crystal_2015}. This extension is non-trivial, since there exist more than one way to choose a unit cell in a periodic system, and therefore representations based on the Coulomb matrix easily becomes non-unique. As pointed out in that paper, the aim when seeking a representation for atomistic systems is to find one that is \emph{(i)} \emph{complete}: incorporates all features of the structural information that are relevant for the underlying problem, but at the same time; \emph{(ii)} \emph{compact}: avoids representation of features irrelevant for the underlying problem (\emph{e.g.}, static rotations); \emph{(iii)} \emph{descriptive}: structural similarity should give representations that are close; and \emph{(iv)} \emph{simple}: low computational effort to compute, and conceptually easy to understand.

The end result of Ref.~\cite{faber_crystal_2015} was three alternative Coulomb matrix inspired representations applicable to periodic crystals. The first one was based on replacing the bare Coulomb interactions in the off-diagonal matrix elements with the corresponding expression for fully periodic systems, \emph{i.e.}, the sum of the total Coulomb interaction energy per unit cell between the infinite periodic lattices of the bare cores of repetitions of two separate atoms in the unit cell. These expressions are evaluated via Ewald sums \cite{ewald_berechnung_1921}. The issue with this expression is that it is somewhat computationally expensive and non-trivial to evaluate correctly. The second generalization of the Coulomb matrix was to duplicate the unit cell a number of times and then use the same expression as for the non-periodic Coulomb matrix, however, with a screened Coulomb interaction (\emph{i.e.}, where the interaction decays exponentially to give a finite reach.) This is very similar to just using the short range term in the Ewald sum. To get an even simpler descriptor, a third expression was invented. It was shown how the Ewald sum can be replaced by an expression that mimics the basic shape and periodicity of the Ewald expression, but which still remains on a simple closed form that is easy to evaluate. This expression was named the `sine' or `sinusoidal' descriptor, because of how it reproduced the periodicity over the unit cell via a sine function.

The three alternative extensions of the Coulomb matrix to periodic systems were tested on a dataset that is now known as FLAA (from the authors' initials). It consists of structures with up to 25 atoms that were randomly selected out of the Materials Project database. In these structures most atomic species occur, in proportions roughly similar to their occurrence in structures published in the literature and extracted into the inorganic crystal structure database (ICSD) \cite{bergerhoff_inorganic_1983, belsky_new_2002} which is the main source of crystal structures for the Materials Project. The conclusion of the 2015 paper \cite{faber_crystal_2015} was that all three alternative extensions of the Coulomb matrix to periodic systems performed approximately equal. The sine descriptor did slightly better than the others, with a 370 meV/atom mean absolute error for predicting formation energies when trained on 3k structures from the FLAA data set.

Two main conclusions follow from the above results. Firstly, the performance of kernel ridge regression-based machines for atomistic systems does not appear to be particularly sensitive to the exact details of how the generalized Coulomb matrix descriptors are constructed, as long as they reasonably well adheres to the aims for a good representation listed above. Secondly, at first glance it may appears as if the performance of the models for molecules far outperforms the corresponding ones for periodic crystals (20~meV/atom vs.\ 0.370~meV/atom). However, the sizes of the chemical space for the two cases are not comparable, and arguably the one used for crystals in Ref.~\cite{faber_crystal_2015} is far larger.

\subsection{Crystal-structure specific models}
To demonstrate that these types of models are capable of reaching a level of accuracy directly useful for applications if one restricts the chemical space, the same authors investigated in 2016 a machine learning model operating on such a smaller space \cite{faber_machine_2016}. This work considered all substitutions of main group elements into four sites of one specific quaternary crystal structure, the elpasolite. This structure was selected because it is the quaternary crystal most frequently occurring with different substitutions in the ICSD database, indicating that this structure can accommodate many types of bonds and thus to be rewarding to characterize fully. High-throughput DFT calculations using the \emph{httk} framework were used to produce data for ca 10k substitutions of elements into the elpasolite crystal structure out a total of 2 million possibilities. Furthermore, a subset of 12 main group elements was selected to give a reduced chemical space of 12k possible substitutions, which were run exhaustively. 

A subsitution into a fixed crystal structure can be uniquely specified by giving which chemical species are at which atomic site in the structure. Hence, the 2016 paper used a very straightforward representation of, essentially, a 2x4 matrix that specified the row and column in the periodic table of the atom species at each of the four sites in the elpasolite structure. This leaves out the precise structural information of the system from the descriptor. \emph{i.e.}, the bond lengths between the atoms. The 2x4 matrix descriptor should be understood to techincally refer to the system relaxed while confined to the elpasolite crystal structure.

\begin{figure}[t]
  \includegraphics[width=\linewidth]{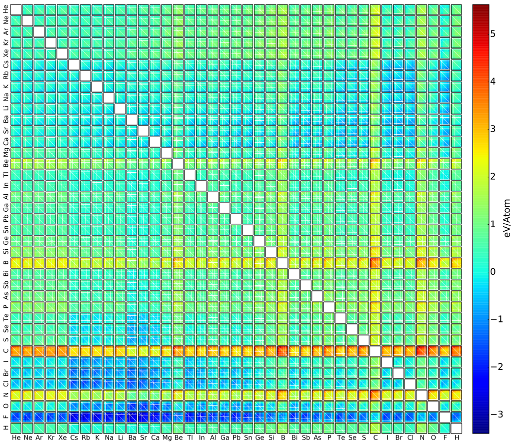}
  \caption{Color matrix of the 2 million elpasolite energies predicted with the crystal-structure specific machine learning model of Faber \emph{et al.} \cite{faber_machine_2016}. The $x$- and $y$-axes specify which atomic species sits on two of the four sites in the crystal structure. At those coordinates one finds a miniature diagram over the species at the remaining two sites. Every pixel in the miniature diagram shows a formation energy of the corresponding composition of four atomic species. The figure is reproduced from the original paper and is licensed under the Creative Commons Attribution 3.0 License.\label{fig:petit} \emph{(Note: this figure has been significantly reduced in quality for online publication, please visit https://doi.org/10.1103/PhysRevLett.117.135502 for a high-quality version.)}}
\end{figure}

A kernel ridge regression machine learning model was trained using this descriptor on formation energies for structures in the elpasolite data set, and it was shown that \emph{(i)} by training on a sufficiently large subset of the exhaustive 12k data set, the model can reach essentially any level of accuracy for predictions of structures outside the training set, at least below $<$ 10 meV/atom which is significantly less than the errors in the DFT data. (See the discussions of accuracy of DFT formation energies in Sec.~\ref{sec:materials_properties}.) This shows that the performance of this machine learning model is merely a question of having a large enough training set; \emph{(ii)} when training on data in the larger chemical space of 2 million possible substitutions of main group elements into the elpasolite structure, it was sufficient to train on about 10k structures to reach roughly the accuracy of the DFT calculations, 100 meV/atom. This result means that the machine learning model was capable of producing DFT-quality formation energies with a net $\times$200 speedup, including all the time used to produce the training data. The resulting 2 million formation energies are illustrated in Fig.~\ref{fig:petit} reproduced from the original paper.

Furthermore, the 2016 paper also demonstrated a practical use of the large set of predicted formation energies. Phase diagrams were created for most of the elpasolite systems by using information about competing compositions from the Materials Project using the pymatgen python library (some systems were outright dismissed on grounds of containing rare-gas elements). From these phase diagrams a number of candidates for thermodynamically stable materials were obtained by identifying compositions with a predicted formation energy on the convex hull. These candidates were validated by DFT calculations and 90 systems were confirmed to be thermodynamically stable within this level of theory. However, the compounds that passed validation only constituted a small fraction of the candidates. As explained in the paper, the reason is that the process of identifying structures on the convex hull is a screening for systems with the lowest formation energies, which are outliers in the full data set. The interpolative nature of machine learning models leads to them being significantly less accurate in predicting properties of outlier systems. Nevertheless, even with this limitation, the scheme far reduced the number of DFT calculations needed to identify thermodynamically stable elpasolite systems compared to just obtaining all formation energies from DFT calculations. The net result was a $\times$11 speedup, including the full time spent both on the training set and the calculations used to validate the materials picked out as candidates for stability.

Hence, the crystal-structure-specific machine was demonstrated to be very successful for generating large amounts of formation energy data which is useful for greatly accelerating predictions of stable compounds in a considered crystal structure. The predictions allow extending the available data in materials genome-type databases. The structures identified as stable in the work discussed above are now available (with some singluar exceptions) via the Materials Project and, \emph{e.g.}, enters the predictions of convex hulls for user-generated phase diagrams via their online service, thus contributing to the accuracy of those predictions.

\subsection{Models for predicting composition phase diagrams, crystal structures}
The success of the crystal-structure-specific machine notwithstanding, in a more typical materials design problem, one usually do not seek all the stable chemical compositions within a single crystal structure, but rather the stable crystal structures that can be formed given a set of chemical species. This is, in essence, the crystal structure prediction problem.

In 2016, Tholander, Andersson, Armiento, Tasn\'adi, and Alling \cite{tholander_strong_2016} (TAATA) produced a data set by high-throughout calculations using the \emph{httk} framework. The aim was to seek stable crystal structures in the ternary chemical systems Ti-Zn-N, Zr-Zn-N and Hf-Zn-N for possible use in piezoelectrics. This high-throughput data-set is a good real-world test case to evaluate the possible acceleration of the generation of phase diagrams for identifying stable structures using machine learning models.

The author of this paper and coworkers have since then engaged in a project of trying out new machine learning models on this problem and to develop new ones for it; the progress on this was recently reported on in \emph{e.g.} Ref.~\cite{bratu_machine_2017}. At the present stage, it appears the original Coulomb matrix-based descriptors from Ref.~\cite{faber_crystal_2015} perform similar on this data set as for the original FLAA data set, which is encouraging in establishing the generality of these models. However, the resulting accuracy is not sufficient to be useful for accelerating the production of the phase diagrams. Compared to the FLAA set, the TAATA dataset has much fewer atomic species, but at the same time is comprised of stuctures over a very wide range of formation energies. The origin of the structures in the FLAA set is the Materials Project which, as explained above, are based on structures from the ICSD database. The ICSD primarily indexes materials seen in nature which means most are thermodynamically stable and have comparably low formation energies. This restriction lowers the dimensionality of the chemical space of FLAA relative to that of TAATA.

Other recent machine learning models perform better; \emph{e.g.}, in Ref.~\cite{bratu_machine_2017} it was found that a descriptor by Ward \emph{et al.} that encodes structural information using a Voronoi tessellation reaches a mean absolute error of 0.28 eV/atom for 10k structures from the TAATA dataset \cite{ward_including_2017}. While errors on this level are not small enough to replace the need for DFT calculations with model predictions, one may still be able to use predictions to identify and remove competing structures that are highly unstable and therefore would not influence the phase diagram, thus reducing the number of DFT calculations necessary, giving an overall reduction in the effort of producing the phase diagram. The field moves rapidly forward, and some other interesting recent developments are found in Refs.~\cite{faber_alchemical_2018,huo_unified_2017,schutt_schnet_2018,ye_deep_2018}.

\section{Conclusions and Outlook}
This paper has reviewed several aspects of producing training data by database-driven high-throughput calculations, and the use of this data to train machine learning models with the aim of accelerating materials design. All these aspects are making rapid and encouraging progress. The research-front machine learning methods are now on the edge of producing results that are accurate and reliable enough to accelerate theoretical prediction of thermodynamic stability via the creation of convex hulls; \emph{i.e.}, the crystal prediction problem which arguably is the most important first step for materials design of bulk materials with desired properties. Further progress towards this goal, and for predicting other properties, is continuously being made. Looking forward, two crucial points can be raised: \emph{(i)} further development of general machine learning models for atomistic systems with improved accuracy and a reduced need for training data is needed; but how far can that development go before it hits a fundamental wall where not enough information about the underlying physics is present in the data?; \emph{(ii)} the rapid development of machine learning models will drive a need for more accurate training data. Will the progress of physics-based computational methods be able to keep up with this need of methods with improved accuracy but low enough computational effort to be useful in high-throughput?; or will the lack of a sufficient amount of high quality training data become a major bottleneck for further progress? Future research needs to target both these areas.

\begin{acknowledgement}
  The author thanks Anatole von Lilienfeld and Felix Faber for many insightful discussions on topics in the overlap of machine learning and materials design. Joel Davidsson is acknowledged for help with supervising the master's thesis discussed in the text as Ref.~\cite{bratu_machine_2017}. The author acknowledges support from the Swedish e-Science Centre (SeRC), Swedish Research Council (VR) Grants No. 2016-04810, and the Centre in Nano science and Nanotechnology (CeNano) at Linköping University. Some of the discussed computations were performed on resources provided by the Swedish National Infrastructure for Computing (SNIC) at the National Supercomputer Centre (NSC) at Linköping University.
\end{acknowledgement}
 
%
%
%

\bibliographystyle{spphys}
\bibliography{refs.bib}

\end{document}